# A Discreet Wearable IoT Sensor for Continuous Transdermal Alcohol Monitoring – Challenges and Opportunities

Baichen Li,[†] Scott R. Downen,[†] Quan Dong,[†] Nam Tran,[†] Maxine LeSaux,[‡] Andrew C. Meltzer,[‡] Zhenyu Li[*,†]

[†] Department of Biomedical Engineering, The George Washington University, 800 22nd ST NW, Washington, DC 20052.
[‡] Department of Emergency Medicine, The George Washington University, Washington, DC.

**ABSTRACT:** Non-invasive continuous alcohol monitoring has potential applications in both population research and in clinical management of acute alcohol intoxication or chronic alcoholism. Current wearable monitors based on transdermal alcohol content (TAC) sensing are relatively bulky and have limited quantification accuracy. Here we describe the development of a discreet wearable transdermal alcohol (TAC) sensor in the form of a wristband or armband. This novel sensor can detect vapor-phase alcohol in perspiration from 0.09 ppm (equivalent to 0.09 mg/dL sweat alcohol concentration at 25 °C under Henry's Law equilibrium) to over 500 ppm at one-minute time resolution. The TAC sensor is powered by a 110 mAh lithium battery that lasts for over 7 days. In addition, the sensor can function as a medical "internet-of-things" (IoT) device by connecting to an Android smartphone gateway via Bluetooth Low Energy (BLE) and upload data to a cloud informatics system. Such wearable IoT sensors may enable large-scale alcohol-related research and personalized management. We also present evidence suggesting a hypothesis that perspiration rate is the dominant factor leading to TAC measurement variabilities, which may inform more reproducible and accurate TAC sensor designs in the future.

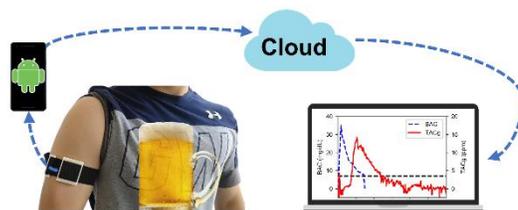

Alcohol consumption is common in the US and around the world with more than half of the US adult population having drank alcohol in the past 30 days.[1] After intake, alcohol (ethanol) molecules pass through the blood-brain barrier and may cause impairment of a person's mental and physical abilities known as intoxication.[2] The annual economic cost to the US due to alcohol is estimated at $44 billion.[3] In addition, over 10,000 deaths are caused by alcohol-impaired driving every year despite comprehensive social education and awareness campaigns. In general, the use of alcohol is associated with myriad negative impacts on disease processes and clinical outcomes.[4,5]

In general, self-reported data or point measurements are required to conduct alcohol-related research and make clinically relevant decisions.[6,7] However, human error and biases inherent in self-reporting may affect the accuracy of collected data.[7–9] Further, collection of objective measurements with high temporal resolution is challenging via self-reporting. A single blood alcohol concentration (BAC) or a single breathalyzed alcohol reading does not give information regarding the rate of metabolism or absorption. To address these challenges, it is desirable to develop a non-invasive and unobtrusive continuous blood alcohol concentration (BAC) monitoring device. Since about 1% of alcohol ingested escapes through perspiration,[10] detecting alcohol released from the skin through a method known as transdermal alcohol content (TAC) sensing and deriving BAC values from these measurements is a promising way to achieve this goal.

There are two wearable TAC sensors that are commercially available. The SCRAM™ and WrisTAS™ are both based on electrochemical transducers. The SCRAM™ focuses on law-enforcement applications, whereas the WrisTAS™ sensors are designed for research. Several laboratory and in-field studies have been conducted to evaluate the performances of both devices.[11–16] However, the data has not conclusively established whether accurate BAC values can be calculated from the TAC data. In addition, the bulky design of the two devices mean that they are not conducive to being worn over an extended period of use. More recently, a third alcohol monitoring device, BACtrack Skyn, is being developed in an unobtrusive wristband form. However, currently this product is not available on the market and little technical information has been released by the manufacturer.

Several research groups have been developing liquid-phase sweat biosensors for alcohol monitoring.[17–21] Liquid-phase sweat analysis differ from TAC sensors in that liquid-phase technologies detect alcohol content in liquid sweat while TAC sensing uses insensible perspiration. Liquid-phase sensing is limited by the challenges in continuous sampling of liquid sweat. As a result, no such devices have been adopted into practical applications such as alcohol-related research and law enforcement.

In this work, we demonstrate a self-contained discreet wearable sensor to continuously monitor vapor-phase alcohol concentration in perspiration. One important design goal of the sensor was to utilize state-of-the-art microelectronics and fuel-cell sensors to enable a wristband-sized form factor, which makes it more user-friendly and nonobtrusive. Furthermore, the wearable sensor is connected to the cloud as an IoT device through an Android smartphone gateway via Bluetooth Low Energy (BLE).[22] By capitalizing on the ubiquity of smartphones and the advancements in cloud systems, the proposed wearable IoT system has the potential to enable large-scale alcohol-related research and management.

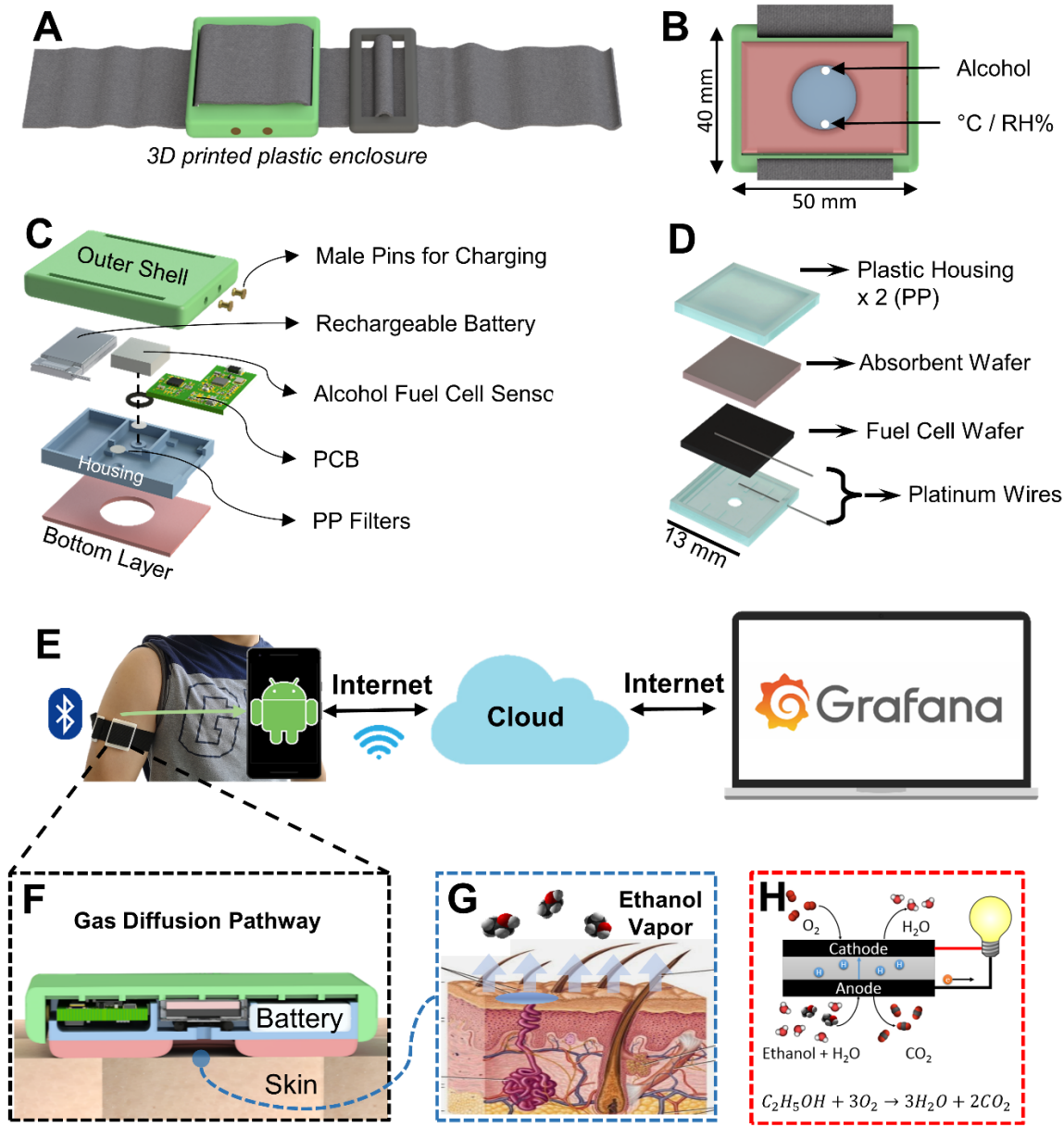

**Figure 1.** Design of the wearable IoT TAC sensor and its working principles. (A) 3D illustration of a fully assembled wearable TAC sensor. (B) Bottom view of the sensor. (C) Exploded view of the internal structure of the sensor. (D) Exploded view of the custom packaging of the fuel cell alcohol sensor. (E) Overview of the cloud-based IoT system including: the wearable sensor, an Android smartphone as the gateway, a cloud informatics system (e.g. AWS) and a web-based user interface. (F) Cross-section view of the wearable sensor, showing the alcohol vapor diffusion pathway. (G) Illustration of the perspiration of ethanol molecules through the skin. (H) Illustration of the working principle of the fuel cell sensor, which converts the presence of ethanol to electrical current.

One important observation in this work is that the vapor-phase TAC sensors showed significant intrapersonal and interpersonal variabilities under nearly identical BAC conditions. To our knowledge, these phenomena have not been sufficiently studied in previous literature. However, for such TAC sensors to be useful as a quantification tool, it is necessary to have reproducible TAC measurements for a given BAC profile. Based on our results, we hypothesize that perspiration rate is the dominant factor leading to the TAC measurement variability. If this hypothesis is correct, we predict more reproducible TAC measurements can be obtained by compensating for the effects caused by changes in perspiration rate.

In the following sections we describe the design, fabrication and testing of our wearable IoT alcohol sensor system, demonstrate the major challenges we encountered, and discuss in detail possible solutions to the observed intrapersonal and interpersonal TAC measurement variability.

## MATERIALS AND METHODS

The wearable IoT alcohol sensor system consists of three sub-systems: (1) a self-contained wireless wearable fuel-cell sensor for transdermal alcohol monitoring, (2) an Android-based smartphone serving as the gateway device to bridge between the sensor and the cloud, and (3) a cloud system deployed on Amazon Web Services (AWS) for data storage and analytics. (Figure 1.E)

**Principle of Transdermal Alcohol Sensing.** After alcohol consumption, ethanol is absorbed by the stomach and small intestine through passive diffusion, further diffusing into the surrounding capillaries. The alcohol molecules then distribute throughout the blood circulatory system, thus reaching other organs in the body, including the skin. Alcohol molecules can escape from the skin through perspiration (Figure 1.G), a process critical to TAC sensing. It has been estimated that approximately 1% of ingested alcohol is excreted through the skin.[10] Following this excretion, the sweat is continuously evaporating, turning a portion of the liquid into gas phase, thus allowing some of the ethanol molecules to diffuse freely in the air. It is through this process that the gas-phase alcohol content can be measured by an alcohol gas sensor.

For clinical purposes, blood alcohol content (BAC) is the target metric to be determined. With a well-controlled environment and, more importantly, stable temperature and local sweating rate, the amount of alcohol vapor released from skin should strongly correlate with the BAC levels. This BAC-to-TAC transformation process can be characterized, and if this process can be described as an invertible function then the BAC levels can in principle be derived from the measurements of a transdermal alcohol sensor.

In our design, the physical quantity measured is the concentration of alcohol vapor in contact with the sensor, converted to parts-per-million (ppm). To avoid any ambiguity in "TAC" between alcohol present in vapor or liquid form, we define a variable specific to the TAC in gas phase: $\boldsymbol{TAC_g}$.

**Fuel Cell-based Alcohol Sensor.** A two-lead electrochemical fuel cell alcohol sensor was selected as the main transducer (Dart Sensors, SKU: 2-EA11). The fuel cell wafer has a sandwich structure: a thin porous substrate in the middle containing aqueous acidic electrolyte for protons to pass through, and platinum-based catalysts coated on both sides of the substrate to form two electrodes,[23] enabling the electrochemical reactions (Figure 1.H).

Like many other direct-ethanol fuel cells (DEFC), this fuel cell sensor catalyzes the reduction-oxidation (redox) reactions between ethanol molecules and oxygen, producing water and carbon dioxide. The half-reactions occurring at the two electrodes of a typical DEFC can be expressed as the following:[24,25]

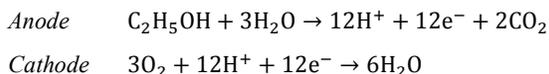

*Anode*   $C_2H_5OH + 3H_2O \rightarrow 12H^+ + 12e^- + 2CO_2$

*Cathode*   $3O_2 + 12H^+ + 12e^- \rightarrow 6H_2O$

Shorting the anode and cathode electrodes externally provides the generated electrons (anode) with a pathway to reach the other side (cathode), as the protons travel through the electrolyte. Thus, electrical current is generated as a product of the overall redox reaction:

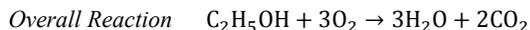

*Overall Reaction*   $C_2H_5OH + 3O_2 \rightarrow 3H_2O + 2CO_2$

According to the manufacturer, the fuel cell sensor generates roughly one electron per ethanol molecule consumed, which is smaller than the theoretical number. Previous studies reveal that the reactions happening in the two electrodes are more complicated than the half-reaction equations, including multiple possible pathways.[26] One hypothetical explanation is that most of the ethanol molecules are first adsorbed on the platinum, generating one electron for each adsorption. The rest of the ethanol molecules would be then oxidized non-electrochemically.

Nonetheless, the sensitivity of the fuel cell sensor is sufficient for detecting transdermal alcohol content with proper mechanical designs, confining the released alcohol molecules in a chamber close to the sensor surface.

As a by-product of the electrochemical reaction, water molecules will be generated at the cathode of the fuel cell. To avoid excessive water accumulation, an absorbent wafer (pink, Figure 1.D) needs to be placed against the cathode electrode.

The dimensions of the original sensor package are approximately $27\ mm \times 25\ mm \times 6\ mm$, while the fuel cell wafer and the absorbent wafer are only $11\ mm \times 11\ mm \times 1\ mm$ each. in order to minimize the size of our overall sensor, we replaced the original package with a custom plastic housing (Figure 1.D).

The parts of the custom plastic housing were fabricated from a 1/4-inch thick polypropylene (PP) sheet (McMaster-Carr, 8742K135) using a CNC milling machine (Roland, MDX-50). Two platinum wires were heated and inserted into the housing bottom using a heated soldering iron to form electrical connections to the inputs of the external amplifier circuits. A fuel cell wafer and an absorbent wafer were extracted from the original package) and placed in the sensor housing with the platinum wires contacting each side of the fuel cell wafer. Finally, the top part of the housing was mounted onto the bottom piece to complete a full sensor assembly (Figure 1.D).

**Device Enclosure Design.** Figure 1.A depicts a 3D illustration of a fully assembled wearable transdermal alcohol sensor, mounted on a 1.5-inch wide elastic spool with a tri-glide slide fixed on one end. The other end of the elastic spool can be inserted into the slide and held in place by friction.

The enclosure of the sensor (Figure 1.C) was designed to hold all components in place and protect them from being damaged, with an overall size of $40mm \times 50mm \times 10mm$ (Figure 1.B). It consists of three separate parts: an outer shell (shown in green), a housing (shown in blue), and a bottom layer (shown in red).

Two venting ports in the housing (Figure 1.B) were designed to provide *independent* diffusion pathways for gas molecules to reach the fuel cell- sensor and a temperature and humidity sensor, separately. This reduces the sensor response time, as the effective dead volume is greatly decreased.

The bottom layer of the enclosure has a thickness of $2mm$ and has a $20mm$ hole in the center (Figure 1.B and C). After the device is put on the body, a small chamber (Figure 1.F) is formed between the skin and the sensor, trapping gas molecules released through perspiration (Figure 1.G). A margin greater than $10mm$ was kept around the hole to ensure a reasonably good seal between the skin and the device and minimizing the chance of gas leakage.

Prototypes of the enclosures were fabricated using a 3D printer (Zortrax M200) and ABS plastic blend filament (Zortrax Z-ULTRAT). After the parts were printed, the bottom layer of the enclosures was glued onto the housing using Gorilla-brand Super Glue and kept in a well-ventilated area for a week.

On completion of the steps above, the enclosures are ready to be assembled with other components of the wearable sensor, as illustrated in Figure 1.C.

**Low-Power Data Acquisition System.** A low-power embedded system was designed based on an ultra-low power BLE microcontroller (Texas Instruments, CC2650RSM) to perform data acquisition, storage, and transmission (Figure 2.A).

A programmable analog front-end (AFE) potentiostat chip (Texas Instruments, LMP91000) with an internal transimpedance amplifier (TIA) was utilized in the system to interface with the fuel cell sensor, converting the sensor output current to voltage. This amplified output signal was directly connected to an analog pin of the microcontroller and digitized by the built-in 12-bit analog-to-digital converter (ADC). Various parameters of the potentiostat, including the bias potential and the amplification gain, can be configured through an inter-integrated circuit ($I^2C$) interface.

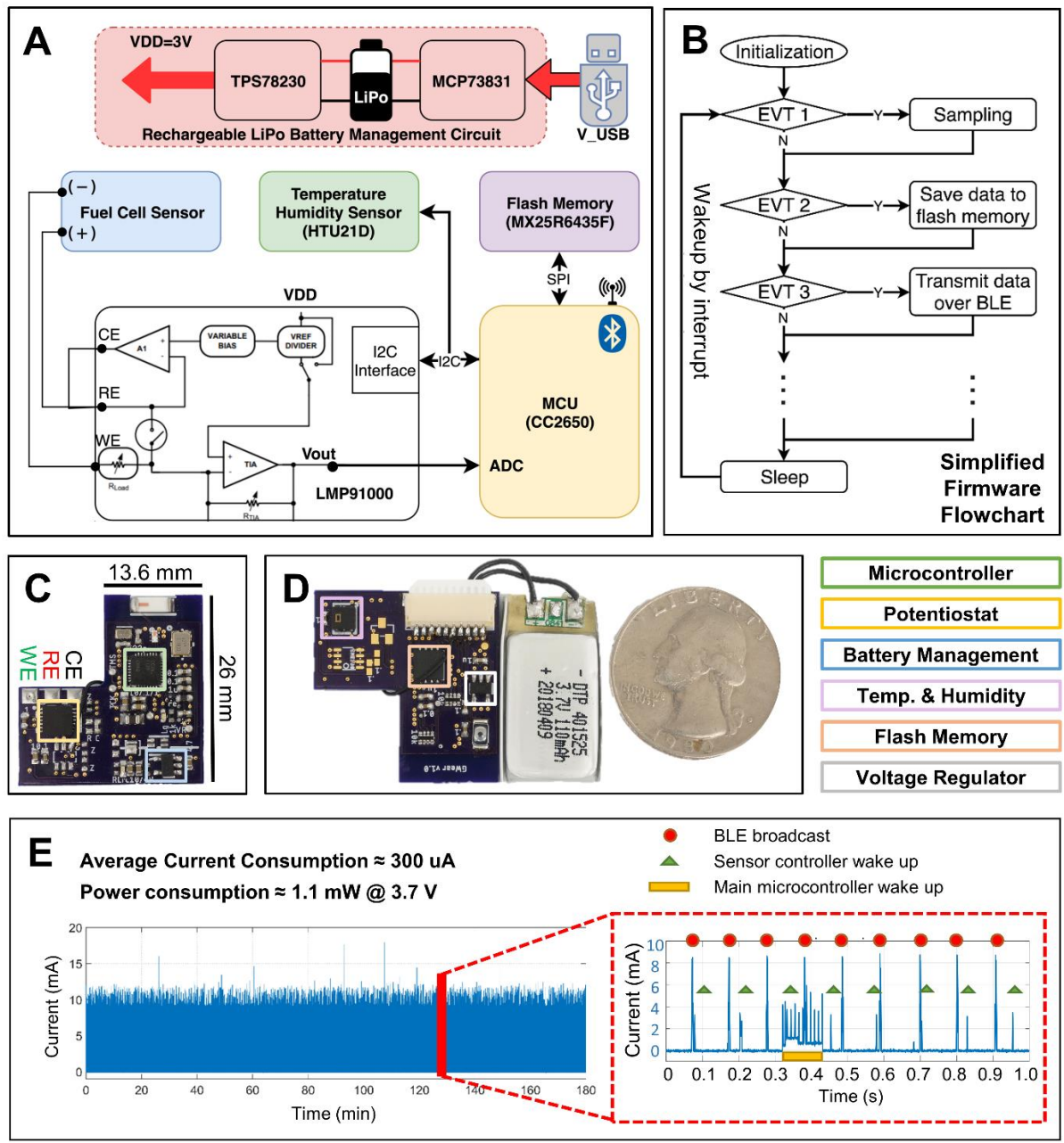

**Figure 2.** Design, fabrication, and characterization of the miniature low-power wireless data acquisition system. (A) Block diagram of the design of the electronics system. (B) Simplified flowchart of the firmware. (C) Top view of an assembled printed circuit board (PCB). (D) Bottom view of the assembled PCB with a lithium battery connected. (E) Average power consumption of the electronics system with the firmware running continuously.

To monitor the environmental conditions in the sweat vapor collection chamber above the skin, a calibrated digital temperature and humidity sensor (HTU21D) was included in this design, which also communicates using the $I^2C$ interface. The reported accuracies of these sensors are ± 0.3 °C and ± 2 %, respectively.

In addition, a low-power flash memory device with the capacity of 8 megabytes (MX25R6435F) was incorporated for general-purpose data storage which communicates with the microcontroller via a serial peripheral interface (SPI).

The whole electronic system was powered by a 110 mAh 3.7V rechargeable Li-ion polymer battery (DTP401525) regulated by a low dropout 3V linear regulator (TPS78230). A dedicated charge management controller (MCP73831) was employed, bridging the external power supply and the battery to enable recharging with a standard 5V USB power supply.

A 4-layer printed circuit board (PCB) layout was designed using Autodesk EAGLE PCB design software (version 8.4.0), consisting of an L-shape geometry (Figure 2.C, D) that allowed us to pack all components into a smaller space. In this layout, a 10-pin, 1.0 mm pitch JST connector was utilized as the interface to access various signal lines in the system.

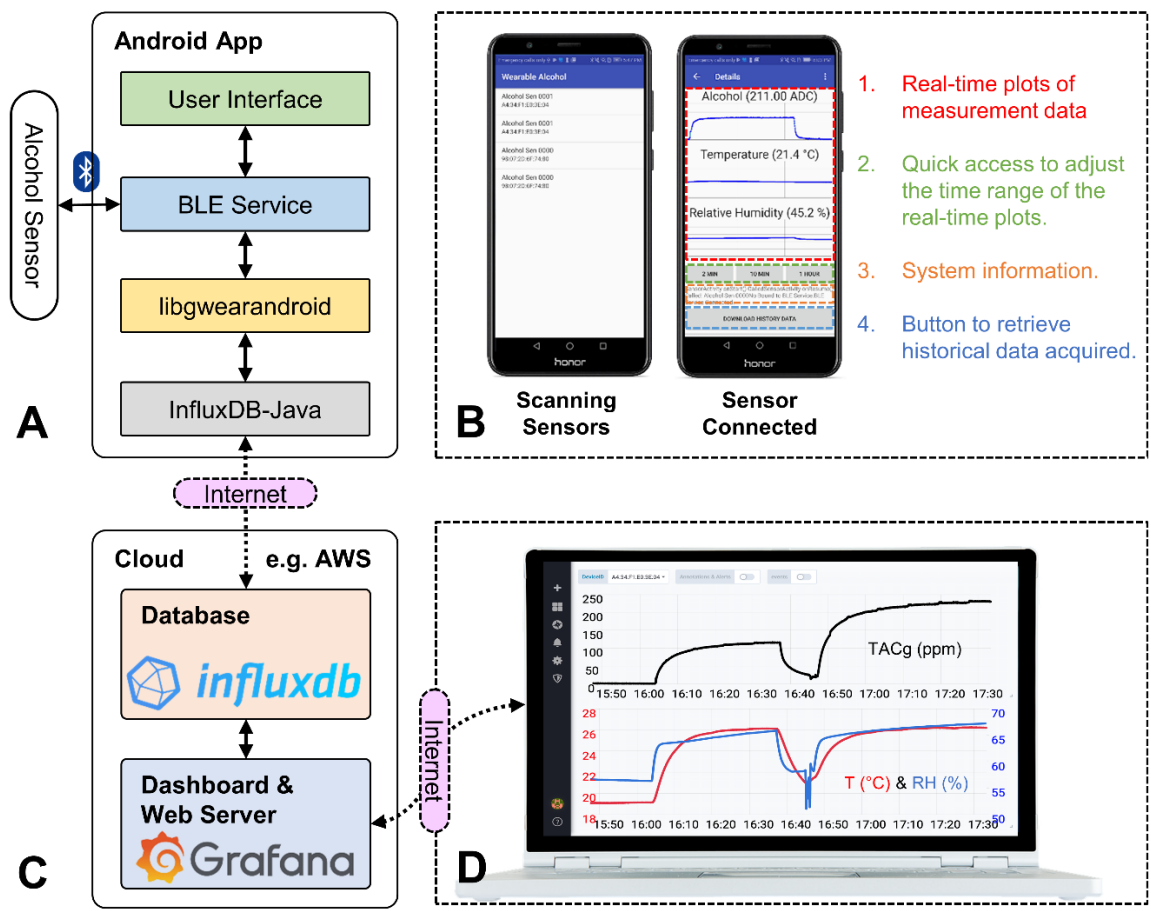

**Figure 3.** The Android app and the Cloud IoT Informatics System. (A) Block diagram of the Android app, consists of (i) a GUI activity, (ii) a BLE service handles the data communication between the smartphone and the wearable alcohol sensor, and (iii) a custom Android library receives the sensor data stream from the BLE service, organizes the data in JSON format, and sends the data to the cloud database using the InfluxDB-java library. (B) Design of the graphical user interface (GUI) of the Android app. (C) Block diagram of the cloud IoT informatics system, including an open-source time-series database (InfluxDB) and a dashboard software with built-in web server (Grafana), hosted on a cloud platform (AWS). (D) A web-based dashboard showing a sample dataset (alcohol, temperature, humidity) acquired from a prototype sensor.

A batch of PCBs was ordered from a commercial PCB manufacturing service (OSHPARK), and the electronic components were populated manually. Afterwards, the fabricated circuit boards were cleaned with the following procedure: (1) immerse in 70 % ethanol solutions for 10 minutes, (2) brush under running pure water multiple times using anti-static brushes, and (3) air-dry for 30 minutes.

The firmware image was loaded into the microcontrollers via the JST connector, and a set of tests were performed to validate the electrical performance of the prototypes. The power consumption of a fully functional circuit is about 1.1mW at 3.7V supply (Figure 2.E), permitting the 110 mAh lithium battery to last for over a week on a full charge.

Once the circuit boards were fully tested, a silicone modified conformal coating (MG Chemicals 422B) was carefully applied on both sides of the PCB to protect it from corrosion. The coated circuits were then kept in a well-ventilated area for a week to let the solvents fully evaporate. This coating is critical for the reliability of the electronic system, as the circuits will be working in a high humidity environment.

**Low-Power Event-Driven Firmware.** A custom firmware was developed using development tools provided by Texas Instruments (TI), including Code Composer Studio (version 8), TI's ARM compiler (v16.9.4.LTS), Sensor Controller Studio (version 2.5.0) and Bluetooth Low Energy software stack (SDK v2.2.2).

A TI RTOS task was implemented to manage and coordinate all peripheral devices asynchronously to accomplish data acquisition, storage, and transmission. It has an event-driven architecture to minimize the power consumption (Figure 2.B).

A Bluetooth Low Energy (BLE) service was implemented for data communication between the wearable sensors and any compatible BLE central devices. This service consists of five major characteristics: two 20-byte characteristics (TX/RX) for general-purpose communication and control based on a custom application layer protocol, and three 4-byte characteristics for spontaneous sensor measurement results (alcohol content, temperature, and relative humidity) in single-precision floating-point format.

After initialization, the electronic system performs data acquisition continuously at a sampling frequency of 1 Hz. The amplified signal of the alcohol sensor is digitized by the ADC of the microcontroller, while the temperature and relative humidity levels are quantified by a calibrated digital sensor (HTU21D). Measurement results are temporarily stored in separate data buffers in the random-access memory (RAM) and written to the value fields of the corresponding BLE characteristics. A software timer, triggered every minute, runs continuously to average the data buffers and store the 1-minute average data into the external flash memory. This flash memory chip is managed as a non-volatile circular first-in first-out (FIFO) buffer, with its storage capacity adequate to hold over 100 days of data.

The LMP91000 potentiostat chip can be configured to select 8 different gain resistances (7 internal and 1 external) for its transimpedance amplifier. An algorithm was implemented to shift the gain automatically based on the digital readings of the ADC, maximizing the dynamic range of the data acquisition system.

**Android App as the Gateway Device.** An Android app was implemented to use a smartphone as the gateway device, bridging the wearable IoT sensor and the cloud system (Figure 3). This app has three major functions: (1) scanning and listing nearby BLE devices with a specific device name, allowing the user to select one to connect, (2) requesting real-time measurement data from a connected device at a frequency of 1 Hz and synchronizing the timestamped data to a dedicated cloud system, and (3) retrieving the historical data stored in the on-board flash memory, timestamping it, and synchronizing to the cloud system. In addition, three buttons were provided as the quick access to adjust the range of the time-series plots.

A simplified block diagram of the app is illustrated in Figure 3.B, which shows the four major software modules. A graphical user interface (Figure 3.A) was designed for the users to interact with the background service (Custom BLE Service) that scans nearby wearable IoT sensors and communicates with a sensor using our custom application layer protocol.

As mentioned above, once a wearable sensor is connected, the real-time measurement data is retrieved at a frequency of 1 Hz. At this point, the user can also download all historical data stored in the flash memory by pressing a button on the GUI. The background service then performs four tasks on every dataset: (1) apply current system timestamp, (2) save a copy in the local storage, (3) send the dataset to the active GUI activity for data display, and (4) transmit formatted data to the cloud system.

A custom library (libgwearandroid) was implemented to compile the retrieved measurement data in JSON format and transmit the formatted data to the cloud system over the Internet using the InfluxDB-java library.

**Cloud System for IoT Data Storage and Analytics.** A prototype cloud system was developed and deployed on the Amazon Elastic Compute Cloud (Amazon EC2) platform for wearable IoT data storage and analytics. An Amazon EC2 instance (t2.micro) was configured and launched in the Amazon Web Services (AWS), managed via remote secure shell (SSH). The instance was configured with 1 virtual CPU, 1 GB memory, and 8 GB SSD storage. Ubuntu Server 18.04 LTS was running on the instance.

An open-source time series database, InfluxDB, was installed to store the timestamped measurement data and miscellaneous information. By default, InfluxDB is listening to a TCP/IP port (8086) for any incoming http request, and incoming data will be inserted into the database through the InfluxDB API.

In addition, an open-source time series data analytics software, Grafana, was utilized for data visualization, permitting the user to view and download measurement data remotely through a web browser. A dashboard (Figure 3.C) was configured to display the measurement data (alcohol, temperature and relative humidity) from a selected wearable IoT sensor in a user-defined time frame. These records can also be downloaded to a local computer for further analysis using data processing software such as MATLAB or Python.

**Data Analysis.** The data entries stored in the wearable sensor share a pre-defined data structure consisting of 16 bytes, including data type (2 bytes), data id (2 bytes), and 3 values (3 x floating-point number). Measurement data is parsed from the binary data retrieved from the sensor, while the timestamps are calculated from the current system timestamp and data id (increase by 1 every minute). This timestamped dataset is stored in a comma separated values (CSV) format for further analysis.

A python program was implemented to perform data analysis, converting the sensor readings to TACg measurements based on the corresponding calibration parameters obtained in the calibration experiments and removing the baseline shift as described below.

**Transdermal Alcohol Sensor Calibration.** Calibration is required to obtain the parameters to convert the ADC readings to the actual alcohol vapor concentrations.

A typical calibration setup is shown in Figure 4.A: an alcohol sensor was taped on the inside of a lid of a sealed mason jar containing 600 mL of an ethanol-water mixture solution of a known concentration, and a nearby smartphone was running our Android-based application to stream real-time sensor data to the cloud. All calibration experiments were conducted in a lab environment at a room temperature of 25 °C. Over time, the gas concentration inside the mason jar reached equilibrium, the value of which was determined by *Henry's law* (Figure 4.B). According to our tests, it generally took 25 to 30 minutes for the sensor reading to plateau, indicating equilibrium had been established.

Six jars of the water-ethanol mixture solutions of various ethanol concentrations were prepared by diluting reagent-grade 99.45% ethanol (Sigma-Aldrich E7023-4L) in pure water. A 50 mL 20 % (w/v) ethanol solution was first prepared by weighing 10 grams of ethanol in a 50 mL centrifuge tube and filling the remainder with pure water. The centrifuge tube was then sealed and attached to a running rotator (VWR 10136-084 Tube Rotator) for 1 hour to assure that the liquids were fully mixed. Afterwards, six mason jars with 600 mL of pure water were prepared, and various volumes of 20% ethanol solution were pipetted into the mason jars to constitute solutions of different concentrations, ranging from 0 to 0.60%.

The alcohol sensor prototypes were placed in each mason jar in sequence. After one calibration routine (six calibration experiments), the data for different concentrations were aligned and data segments at 25-30 minutes were extracted to derive the calibration curve. Figure 4.C shows a sample dataset obtained from a calibration routine. As expected, the readings of the alcohol sensors are linearly proportional to the ambient ethanol gas concentration.

**Sensor Baseline Characterization.** The baselines of the fabricated alcohol sensors were evaluated after the calibration. A calibrated sensor was placed in a lab environment (temperature-controlled room, T≈25°C) for an extended period (> 2 hours) without the presence of alcohol, establishing the baseline of the sensor in a "clean" environment. Afterwards, a male human subject was instructed to wear the sensor on his left upper arm for 2-4 hours in the same environment without ingesting alcohol to observe the behavior of the baseline during on-body deployment.

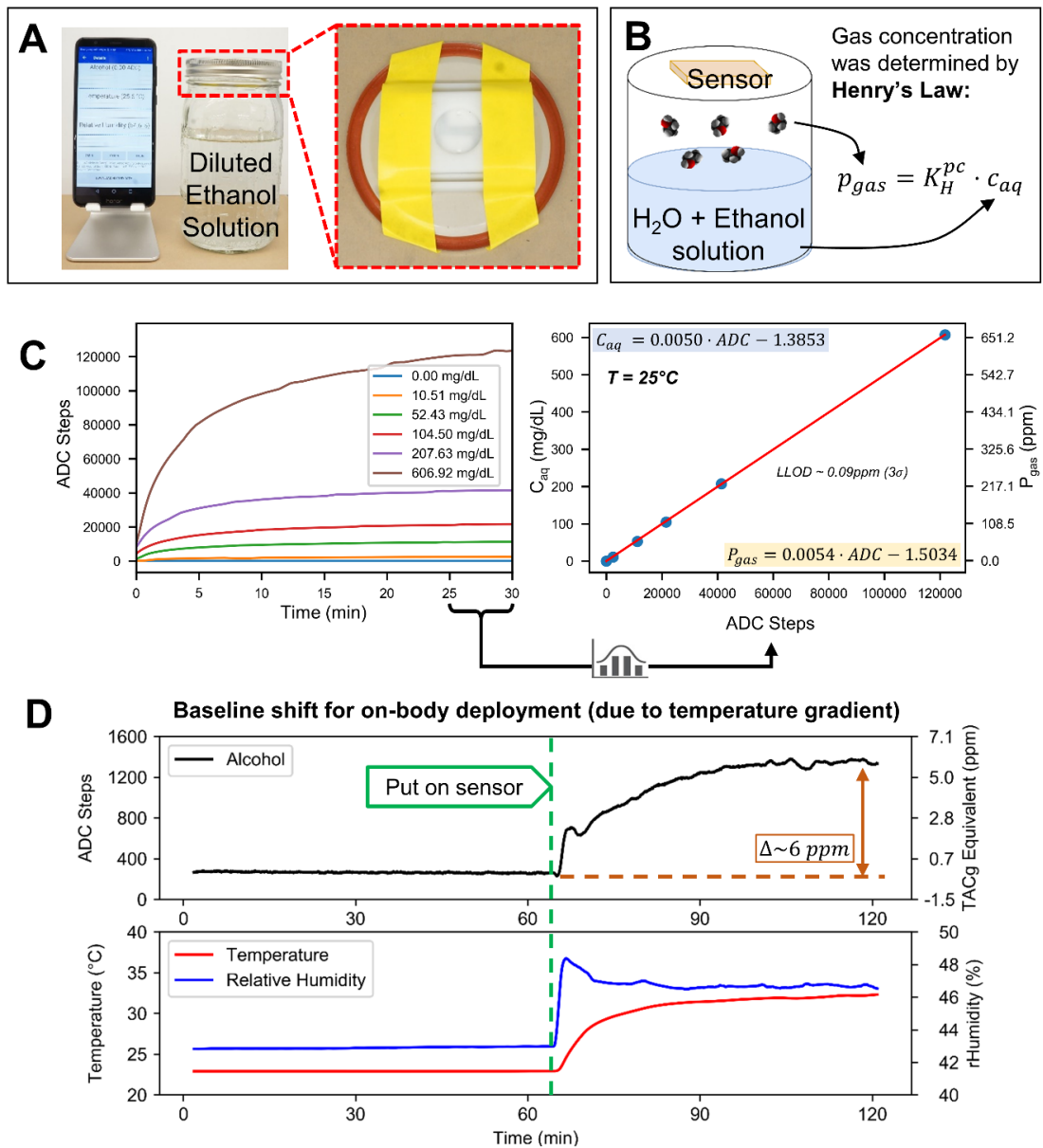

**Figure 4.** Characterization of the TAC sensor. (A) Experimental setup for sensor calibration using known concentration ethanol-water solutions in a sealed mason jar. (B) The vapor-phase ethanol concentration inside the sealed mason jar is determined from the liquid-phase concentration by Henry's law under equilibrium. (C) A sample sensor calibration using six prepared ethanol-water solutions with known concentrations. The sensor calibration curve is shown on the right. (D) Results of an on-body deployment experiment with no alcohol consumption, showing a baseline shift caused by temperature gradient between the human body and the environment. The baseline shift stabilized after approximately 1 hour and was removed in data post-processing.

A sample dataset of the baseline characterization experiment is shown in Figure 4.D. The measurement data of the sensors was very flat before the on-body deployment, with mean ($\mu_{ADC}$) = 265.8 (-0.07 ppm) and standard deviation ($\sigma_{ADC}$) = 5.8 (0.03 ppm). However, after the sensor was placed on the body, changes in sensor readings were observed. The baseline shift of the alcohol sensor reached a plateau in roughly an hour, following the trends of the temperature data. The mean of the alcohol sensor measurement in plateau was increased to 1340.6 (5.74 ppm), and the standard deviation was increased to 23.8 (0.13 ppm).

According to our observations, both the magnitude and the direction of this sensor baseline shift were mainly affected by the temperature gradient across the sensor electrodes, which is generated by the difference between body temperature and the ambient temperature. Currently, the underlying mechanism is not clear to us. However, combining the observations and the nature of heat transfer processes, we suspect that the baseline shift follows a shifted negative exponential function ($\pm A \cdot e^{-Bt} \mp C$). In practice, this baseline shift was approximated using a second-order polynomial and subtracted from the original measurement data.

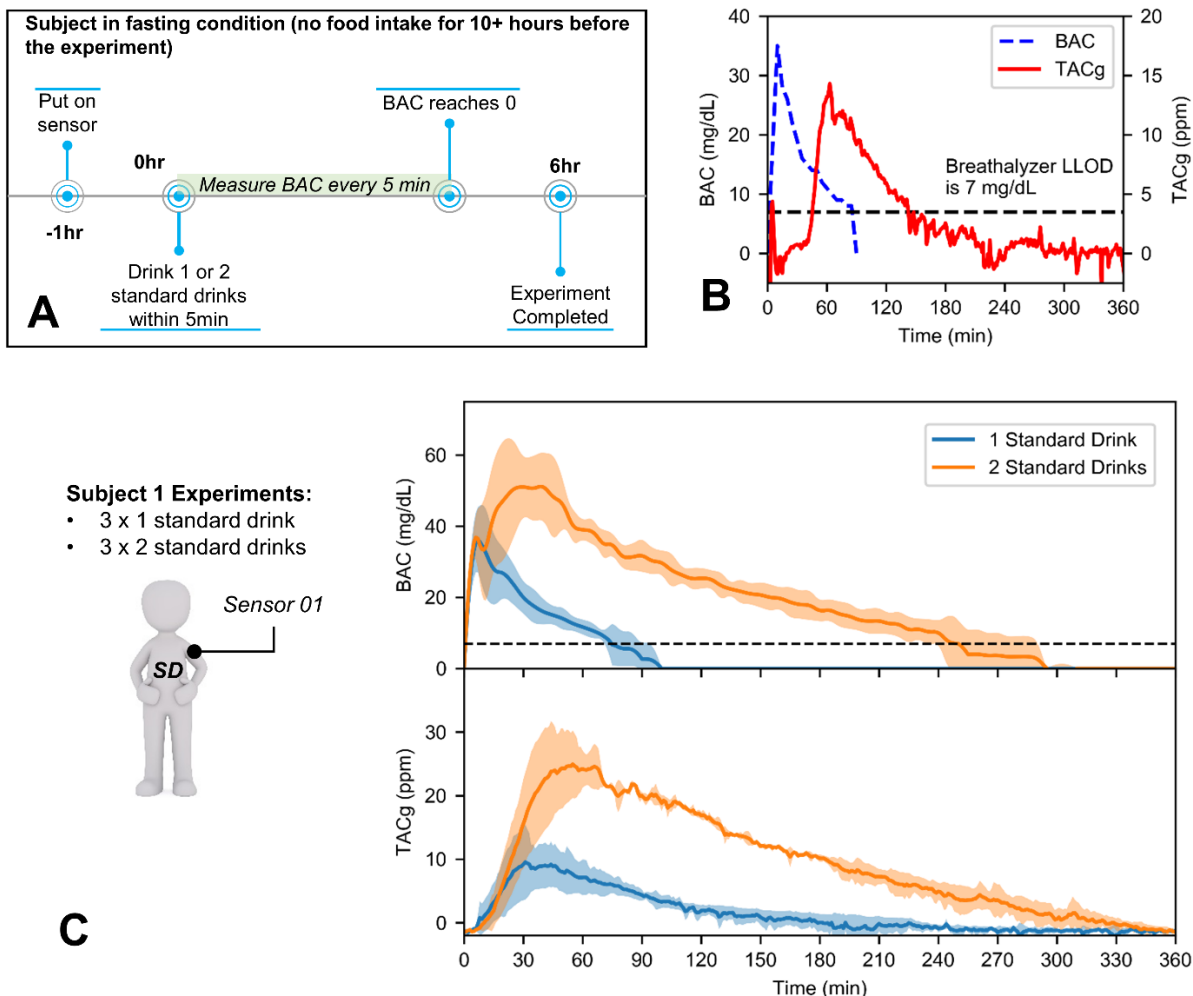

**Figure 5.** Human subject testing under well-controlled environmental and drinking conditions. (A) The protocol of a controlled alcohol consumption experiment with the transdermal alcohol sensor attached on the left upper arm of the subject. (B) The figure of a sample set of BAC and TACg data from an one standard drink experiment, showing that relative to the BAC curve, the TACg is delayed and broadened, consistent with previous studies.[27–31] (C) Plots of the average BAC and TACg curves derived from six experiments conducted by a male human subject, with color bands highlighting the corresponding standard deviations. The results showed both BAC and TACg curves can easily distinguish one and two standard drinks.

**Study Protocol and Sensor Testing with Human Subjects.** The study protocol is depicted in Figure 5.A. The experiments were completed in a lab environment, where the ambient environment is relatively stable (especially temperature), minimizing the influences of the various random variables present in a real-world environment. A 10+ hour fasting subject was instructed to wear the alcohol sensor and wait for an hour for the baseline stabilize. Next, the subject consumed one or two standard drinks (Sake, 15% alcohol content) within five minutes and rinsed his mouth thoroughly three times to eliminate any residual mouth alcohol.[32] A breathalyzer (BACtrack S80) was provided for the subject to measure his/her BAC level every 5 minutes, until it indicated a reading of 0. The subject was asked to avoid food and water intake during this period for consistent alcohol absorption in the body.

Once the BAC level reaches zero, there should be no alcohol left in the digestive system and the subject can ingest anything without alcohol content without negatively impacting the study. Previous studies have shown that it takes much longer for the TAC to vanish.[33] To account for this, the subject was instructed to wear the sensor for 6 hours after drinking. Historical data stored on the sensor were downloaded for analysis after the experiments.

Following the protocols outlined above, six alcohol consumption experiments were conducted by a male subject (subject 1) to validate the performance of the sensor prototypes and study the variability of both BAC and TACg – three with one standard drink and three with two standard drinks. Based on the observations from the initial six trials, additional comparison studies were designed and conducted by adding a second male subject (subject 2).

**Results**

**BAC and TACg data (Figure 5).** Figure 5.B shows the results from the initial human subject study, which include two time-series data of BAC (mg/dL) and calibrated TACg (ppm) measurements. The BAC measurement obtained with a breathalyzer was used as "the gold standard" to compare against its corresponding TACg values. Results agreed with previous studies morphologically.[27–31] However, the absolute TACg measurements (ppm) and the corresponding aqueous alcohol concentrations were generally smaller than the results shown in the previous literature.[27–31] We suspect that the commercial sensor systems utilized in these studies may have some internal calibration mechanisms to better estimate the BAC values.

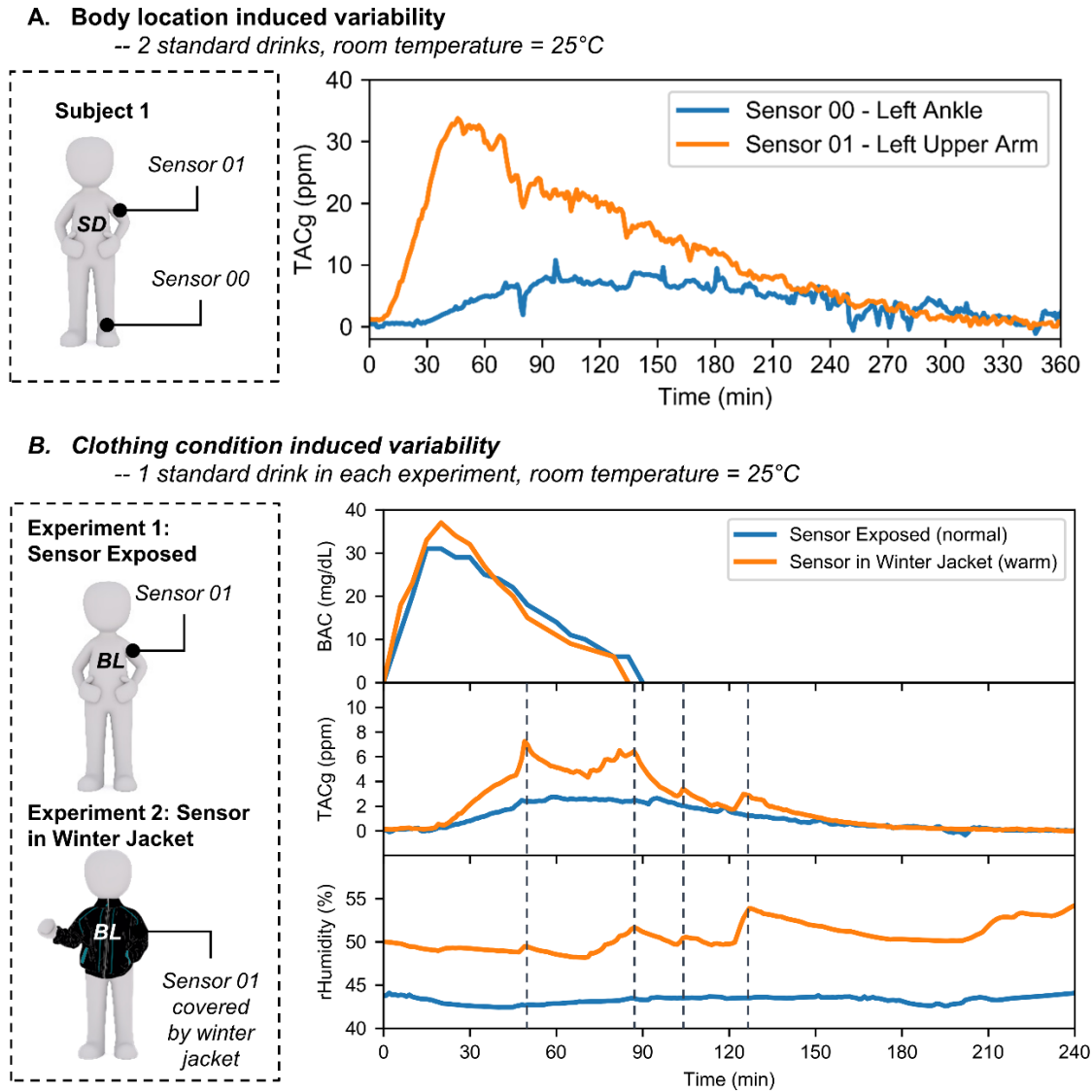

**Figure 6.** Intrapersonal variability of transdermal alcohol sensing. (A) Body location induced variability: two calibrated sensors attached to different parts of the body (left upper arm and left ankle) produced different results in a single drinking experiment. The TACg data obtained from the upper arm was significantly higher than that from the ankle. (B) Clothing condition induced variability: BAC and TACg results of two experiments conducted by a male subject consuming one standard drink each. In the first experiment, the subject was wearing a t-shirt. In the second, the subject was wearing winter jackets to induce sweat. The BAC curves agreed well with each other, while the TACg results showed significant difference. Whenever there is a humidity spike, a TACg spike was also observed (as marked by the dashed vertical lines in the figure).

After completing the initial six human subject trials, the mean values of the BAC and TACg under different doses were plotted in two separate graphs, with color bands depicting the corresponding standard deviations (Figure 5.C). Both BAC and TACg data for different doses can be easily distinguished, which is a positive sign for being able to utilize the TACg data to estimate the BAC levels.

**Intrapersonal Variability of TACg measurements (Figure 6).** To observe potential differences in TACg measurements due to the sensor's position on the body, in some of the experiments the subjects were instructed to wear calibrated sensors on both the left upper arm and the left ankle. Figure 6.A shows a sample set of TACg data, comparing the two alcohol sensor readings. As shown, the TACg measurements were substantially different, which suggests that different locations on the body can produce very different TACg results. One explanation for this observed difference is a potential difference in local perspiration rates.

To demonstrate that perspiration rate will affect the amount of alcohol released through the skin (TACg measurements), two identical experiments were conducted by the second male subject (subject 2), with the only difference being the subject was wearing three layers of thick jackets in the second experiment to induce sweating (Figure 6.B). We used the ratio between the areas under curves (AUC) as a measurement to evaluate the BAC and TACg differences. The ratio of the BAC was 0.98, while that of the TACg was 0.52. It was also found that during the second experiment, spikes in the TACg were correlated with spikes in relative humidity (marked by vertical dashed lines).

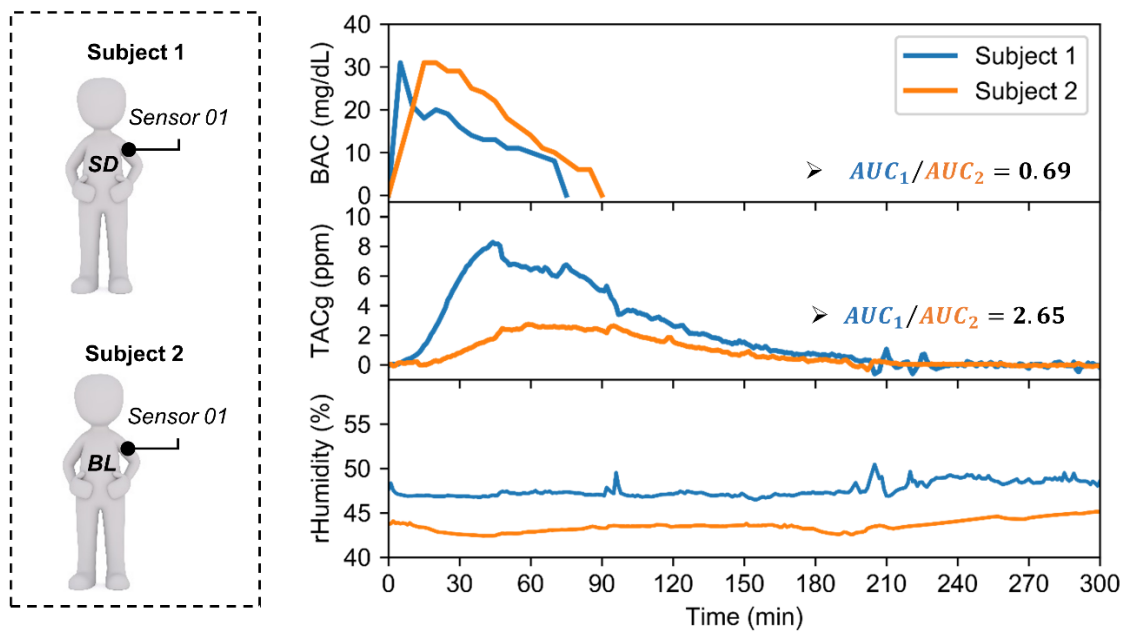

**Figure 7.** Interpersonal variability of transdermal alcohol sensing: two male subjects, each consuming one standard drink, with the same sensor attached to the same location on the body (left upper arm), under similar ambient conditions. The BAC results were slightly different, with the ratio between the areas under curves (AUCs) being 0.69. However, the TACg results were significantly different, with the ratio between AUCs being 2.65. The relative humidity data from subject 1 was also consistently higher than that from subject 2.

However, this correlation was no longer present at the end of the experiment, likely because there was very little alcohol content in the perspiration.

**Interpersonal Variability of TACg Measurements (Figure 7).** It was also observed that TACg measurements from different subjects can be notably different while using the same sensor, sensor location, and amount of alcohol consumed. The comparison of experiments conducted by two male subjects after consuming one standard drink are shown in Figure 7. The AUC ratio of the BAC was 0.69, while the ratio of the TACg was 2.65. This implies that in these two experiments, subject 1 had a lower BAC levels but released more substances through the skin perspiration. This contradiction can also be explained by the interpersonal differences in perspiration rates, as the relative humidity data from subject 1 was consistently higher than that from the subject 2.

## Discussion

Arguably, the ultimate goal for personalized alcohol monitor development is to *achieve a discreet non-invasive continuous sensor capable of measuring BAC levels in real-time*. Despite continuous efforts, this goal remains elusive due to the daunting challenges in performing continuous blood measurements. TACg sensing offers a promising alternative by measuring alcohol in insensible perspiration. In addition, previous theoretical studies suggest that it is possible to derive BAC from TACg data,[34–36] which implicitly assumes that the BAC-to-TACg transformation is invertible, i.e. there is a one-to-one mapping between BAC and TACg. In practice, this dictates that for a given BAC profile the TACg sensor should give a reproducible output, and distinct BAC profiles should lead to distinct TACg curves. However, as described above, we observed significant intrapersonal and interpersonal variabilities in TACg measurements under nearly identical BAC conditions. Our data suggest that intrapersonal variabilities are likely caused by changes in perspiration rate which occur in all TACg sensors. The interpersonal variability issues can potentially be resolved by personalized calibrations, as long as the TACg measurements are reproducible for an individual under a given BAC profile.

It is important to emphasize that for TACg sensors to be useful as a quantification tool, it is necessary to have reproducible TACg measurements for a given BAC profile. Therefore, below we discuss in more detail how to address the measurement variabilities issues, and other challenges and opportunities in improving TACg sensing technology.

**Effect of the Perspiration Rate.** In our experiments, we observed the TACg sensor output has a strong correlation with relative humidity (Figure 6.B). Because the studies were conducted in a climate-controlled environment, the most likely source for local humidity change is perspiration. Thus, we hypothesize that the dominant factor in determining the intrapersonal variability in TACg measurements is an individual's perspiration rate.

As fuel cell sensors operate by consuming vapor-phase ethanol molecules to generate an electrical current, an increase in the ethanol gas concentration in the cell's vicinity will lead to higher sensor readings. However, being the source of transdermal alcohol generation, the overall perspiration rate of an individual can change dramatically, ranging from 20.8 mL/hr to 1.8 L/hr.[37–41] Such variation will result in significantly different gas concentrations in the collection chamber for the same alcohol content in liquid sweat.[*]

We believe that left uncompensated, other TAC sensors employing similar sensing mechanisms will also be susceptible to changes in perspiration rates. This can potentially explain the variabilities and errors reported in previous studies using TAC sensors.[13,42]

---

[*] Except for special cases, such as when the chamber reaches Henry's Law equilibrium.

Future studies are required to test this hypothesis and determine if more reproducible TACg measurements can be obtained by compensating for perspiration rate changes. To quantify perspiration rate continuously, various techniques have been developed by the research community, including using multiple humidity sensors,[43,44] and measuring skin impedance.[45–47]

**Need for a Theoretical Model of Vapor-Phase TAC Sensing.** It is also desirable to establish a theoretical model of the wearable alcohol sensor to gain a deeper understanding of the overall gas-phase transdermal alcohol movement and provide insightful guidance for sensor design and data analysis. From this model, conditions under which BAC levels can be derived from the TACg measurements may be evaluated.

Based on our current understanding, there are five major physical processes involved between BAC and TACg measurements: (1) capillary circulation of the skin,[48] (2) fluid and solute exchange in the skin (diffusion), (3) evaporation of sweat, (4) diffusion of ethanol molecules in the sensor spaces, and (5) fuel cell sensor reactions.

A few intriguing publications on mathematical models for the underlying physiological systems and using computational tools to improve the transdermal alcohol sensing results have been published. Topics covered include the fluid and solute exchange in the human system,[49] the transdermal transportation of alcohol,[50] the microfluidics of sweat glands,[51] the kinetics of transdermal ethanol exchange,[52] and the deconvolution of transdermal ethanol sensor data,[34–36] to name a few. These works provide a good framework for creating a model for the wearable alcohol sensor.

As they stand, transdermal alcohol sensing technologies are not a direct replacement to currently accepted alcohol sensing technologies,[33] such as breathalyzers, blood tests, and infrared analysis. Although promising results have been published to estimate BAC from TAC data, it is still very challenging, if not impossible, to achieve real-time BAC measurement using a transdermal alcohol sensor, mainly due to the time delays between the BAC and the TAC. Instead, it is more likely that a final device will perform continuous TAC measurements and transform the data to time-series BAC results afterwards.

**Effects of Environmental Factors on the Fuel Cell-based Alcohol Sensor.** Although reproducible results can be obtained in a well-controlled environment, significant signal fluctuations were observed when wearing the sensors in an outdoor environment with or without drinking, making it challenging to understand and interpret the data. The most probable influencing factors include temperature fluctuations, interfering gases, sweating rate variations, and motion artifacts.

Baseline shifts due to environmental factors were observed in our experiments, as the fuel cell-based alcohol sensor is sensitive to thermal gradients. Particularly in applications with low-dose alcohol consumption, it is important to remove the thermally related baseline shift from the signal. While in a well-controlled environment, the baseline shift can be approximated and corrected by polynomial fitting, in a real-world setting with less control of ambient temperature no simple solution has been identified to accurately remove the baseline shift. This problem can be tackled with multiple approaches, including using better thermal insulation designs, compensating for the thermal gradient by using multiple temperature sensors.

Moreover, the fuel cell sensor has a transient response to sudden humidity changes. Our testing results showed that a signal spike, lasting for 5-10 minutes, would appear after a sudden change in humidity. The magnitude of the spike was typically smaller than 3ppm, and eventually vanished. More data is needed to determine if this transient response can be reduced or eliminated based on the humidity sensor data.

**Specificity of the Fuel Cell-based Alcohol Sensor.** It is very challenging for low-power miniature gas sensors to be highly specific.[53] The specificity of the fuel cell sensor utilized in this project is mainly determined by the platinum-based catalysts coated on the sensor electrodes, making the sensor only sensitive to a few groups of substances.[54] This fuel cell sensor is most sensitive to various alcohols and aldehydes, which normally are not produced by the human body. However, it is possible for interfering gases to be present in a complicated real-world environment, thus affecting the sensor measurements. Integrating more sensors into the wearable sensor and applying the emerging machine learning techniques may help improve specificity of the wearable sensor.[55–58]

**Gas Pathway Design and Material Selections (Sensor Response Time).** Alcohol vapor and other interfering gases trapped in the alcohol sensor can also affect the baseline and response time of the fuel cell sensors. We found that isolating the gas pathway to the fuel cell sensor from the other internal spaces of the wearable device is an effective method to reduce the time constant of the sensor. Additionally, the prototype PCBs were cleaned thoroughly, and the coated and glued parts were kept in a well-ventilated area for a week for the solvents to fully evaporate, minimizing any potential influences from the solvent residue. For applications monitoring very low concentrations of gases (e.g. in the parts-per-billion range), it is desirable to fabricate the enclosures using metals with proper chemical resistance to minimize gas absorption.

**Potential Applications.** We envision that in the future such a wearable transdermal alcohol sensor capable of estimating BAC data will be useful for the alcohol-related research community, for obtaining objective measurement data at a high temporal resolution. It will also be useful for certain specialized applications where contextual information can be collected, such as emergency rooms, police stations, etc. Additionally, this wearable sensor design and the analysis are not limited to transdermal alcohol sensing. Any volatile substance in the sweat can be detected by integrating a proper transducer into the system.

## Conclusions

In this work, we designed and implemented a wearable IoT sensor for continuous transdermal alcohol monitoring. Wearable sensor prototypes were fabricated and characterized in a laboratory environment and validated by human subject studies with alcohol consumption under well-controlled conditions.

We demonstrated that it is feasible to acquire continuous TAC data (TACg) using our prototype system. Results from repeated alcohol consumption experiments showed that different doses of alcohol consumption can be easily distinguished using the TACg. However, interpersonal and intrapersonal variabilities in TACg measurements were observed in several comparative experiments. Based on our observations, we formulated a hypothesis that *the major influencing factor is the changes in perspiration rate*. If this hypothesis is correct, more reproducible TAC measurements can be derived by correcting for the effects caused by changes in perspiration rate. Theoretical analysis and further validation experiments should be conducted in future studies to test this hypothesis and improve the TAC sensors.

## ASSOCIATED CONTENT

There is no additional supporting information for this version of the manuscript.

## AUTHOR INFORMATION

### Corresponding Author

*E-mail: zhenyu@gwu.edu (Z.L.)

### Funding Sources

National Institute on Alcohol Abuse and Alcoholism of the National Institutes of Health Award Number 1R41AA02464801.

### Notes

The authors declare no competing financial interest.

## ACKNOWLEDGMENT

Research reported in this publication was supported by the National Institute on Alcohol Abuse and Alcoholism of the National Institutes of Health under Award Number 1R41AA02464801. The content is solely the responsibility of the authors and does not necessarily represent the official views of the National Institutes of Health. The authors also thank Dr. Walter King for helpful discussions on the fuel cell sensor technologies.

## ABBREVIATIONS

BAC, blood alcohol content; TAC, transdermal alcohol content; TACg, vapor-phase transdermal alcohol content; IoT, internet of things;